\documentclass{sig-alternate}

\usepackage{balance}
\usepackage{url}
\usepackage{xspace}
\usepackage{listings}
\usepackage{tabularx}
\usepackage{subcaption}
\usepackage{color}
\usepackage[ruled,lined,linesnumbered]{algorithm2e}
\usepackage{placeins}
\usepackage{verbatim}
\usepackage{multirow}

\newcommand{\system}{\emph{HashStash}}

\begin{document}

\title{Revisiting Reuse in Main Memory Database Systems}

\numberofauthors{1}
\author{
\alignauthor
\vspace{-8mm}
\begin{tabular}{cccc}
Kayhan Dursun & Carsten Binnig & Ugur Cetintemel & Tim Kraska
\end{tabular}\\
\vspace{1.5mm}
\begin{tabular}{c}
Brown University \\
\{firstname\_lastname\}@brown.edu
\end{tabular}
}
\maketitle

\begin{abstract}

Reusing intermediates in databases to speed-up analytical query processing has been studied in the past.
Existing solutions typically require intermediate results of individual operators to be materialized into temporary tables to be considered for reuse in subsequent queries. 
However, these approaches are fundamentally ill-suited for use in modern main memory databases. 
The reason is that modern main memory DBMSs are typically limited by the bandwidth of the memory bus, thus query execution is heavily optimized to keep tuples in the CPU caches and registers.
To that end, adding additional materialization operations into a query plan not only add additional traffic to the memory bus but more importantly prevent the important cache- and register-locality opportunities resulting in high performance penalties.

In this paper we study a novel reuse model for intermediates, which caches internal physical data structures materialized during query processing (due to pipeline breakers) and externalizes them so that they become reusable for upcoming operations. 
We focus on hash tables, the most commonly used internal data structure in main memory databases to perform join and aggregation operations. 
As queries arrive, our {\em reuse-aware optimizer} reasons about the reuse opportunities for hash tables, employing cost models that take into account hash table statistics together with the CPU and data movement costs within the cache hierarchy. 
Experimental results, based on our \system{} prototype demonstrate performance gains of $2\times$ for typical analytical workloads with no additional overhead for materializing intermediates.  

\end{abstract}

\vspace{-1.5ex}
\section{Introduction}
\label{sec:introduction}

\textbf{Motivation:} Reusing intermediates in databases to speed-up analytical query processing has been studied in the past ~\cite{monetdb2009,cod2001,vvreuse_13,qpipe2005,qpipe2006,mrshare,mrmqo}.
These solutions typically require intermediate results of individual operators be materialized into temporary tables to be considered for reuse in subsequent queries. 
However, these approaches are fundamentally ill-suited for use in modern main memory databases. 
The reason is that modern main memory DBMSs are typically limited by the bandwidth of the memory bus and query execution is thus heavily optimized to keep tuples in the CPU caches and registers ~\cite{hyper2011,DBLP:reference/db/Ross09,llvm}.

To that end, adding additional materialization operations into a query plan not only add additional traffic to the memory bus but more importantly prevent the important cache- and register-locality, which results in high performance penalties.
Consequently, the benefits of materia\-lization-based reuse techniques heavily depends on the characteristics of the workload: i.e., how much overlap between queries of a given workload exists.
In the worst case, if the overlap is low the extra cost caused by materialization operations might even result in an overall performance degradation for analytical workloads.

The goal of this paper is to revisit "reuse" in the context of modern main memory databases~\cite{hyper2011,hana2013,tupleware2015}. 
The main idea is to leverage internal data structures for reuse that are materialized anyway by pipeline breakers during query execution.
This way, reuse comes for free without any additional materialization costs.
Moreover, as a result reuse becomes more robust towards workloads with different reuse potentials and provides benefits for a wider range of workloads even if the overlap between queries is not that high.

In this paper, we present our system called \system{} that implements reuse of internal data structures.
The focus of this work is on the most common internal data structure, hash tables (HTs), as found in hash-join and hash-aggregate operations. 
We leave other operators and data structures (e.g., trees for sorting) for future work.\\

\vspace{-1.5ex}
\textbf{Contributions:} To the best of our knowledge this is the first paper that studies the reuse of internal data structures for query processing.
As a major contribution, we present a new system called \system{} that extends a classical DBMS architecture to support the reuse of internal hash tables.
The architecture of \system{} supports two reuse models:

\emph{(1) Single-query Reuse}: In this re-use model, users or applications submit a single query to a \system{}-based DBMS just as in normal DBMSs.
However, different from a classical DBMS, a \system{}-based DBMS identifies the best {\em reuse-aware plan} that leverages existing intermediate hash tables.
To support this model, we extend the DBMS architecture by three components: (a) {\em a cache for hash tables} that keeps lineage and statistics information, (b) {\em a reuse-aware optimizer} that uses new operator cost models and enumeration strategies to determine which hash tables should be reused by which operators in order to minimize the total query runtime, and finally (c) {\em a garbage collector} that evicts hash tables from the cache as needed using eviction strategies.

\emph{(2) Multi-query Reuse}: Many analytical applications today execute multiple queries at the same time to show different aspects over the same data set.
In order to support multiple queries that are submitted at the same time, we leverage the concept of shared plans as introduced in \cite{sharedb2014, sharedb2012} and extend them in the following directions:
(a) we develop {\em shared reuse-aware plans}, i.e., shared plans can also re-use the hash tables in~\system{} and
(b) we extend the optimizer in~\system{} to create optimal reuse-aware shared plans for a given batch of queries.

Finally, we evaluate the performance of the \system{} system under workloads with different reuse potentials. 
Our experiments show that \system{} outperforms materia\-lization-based reuse for any of these workloads independent of the reuse potential.\\

\vspace{-1.5ex}
\textbf{Outline:} The rest of this paper is structured as follows.
Section \ref{sec:arch} gives an overview of our suggested \system{}-based  architecture to support single-query and multi-query reuse.
Section \ref{sec:single} and Section \ref{sec:multi} then present the details for each of these reuse cases and discuss novel optimization strategies to support them.
Afterwards, Section \ref{sec:garbage} discusses how garbage collection works in \system{}. 
Section \ref{sec:eval} presents our evaluation of our \system{} prototype. 
Finally, Section \ref{sec:related} discusses related work and Section \ref{sec:concl} concludes with a summary and outlines potential future work.
\vspace{-1.5ex}
\section{HashStash Overview}
\label{sec:arch}

\begin{figure}
\centering
\includegraphics[width=0.47\textwidth]{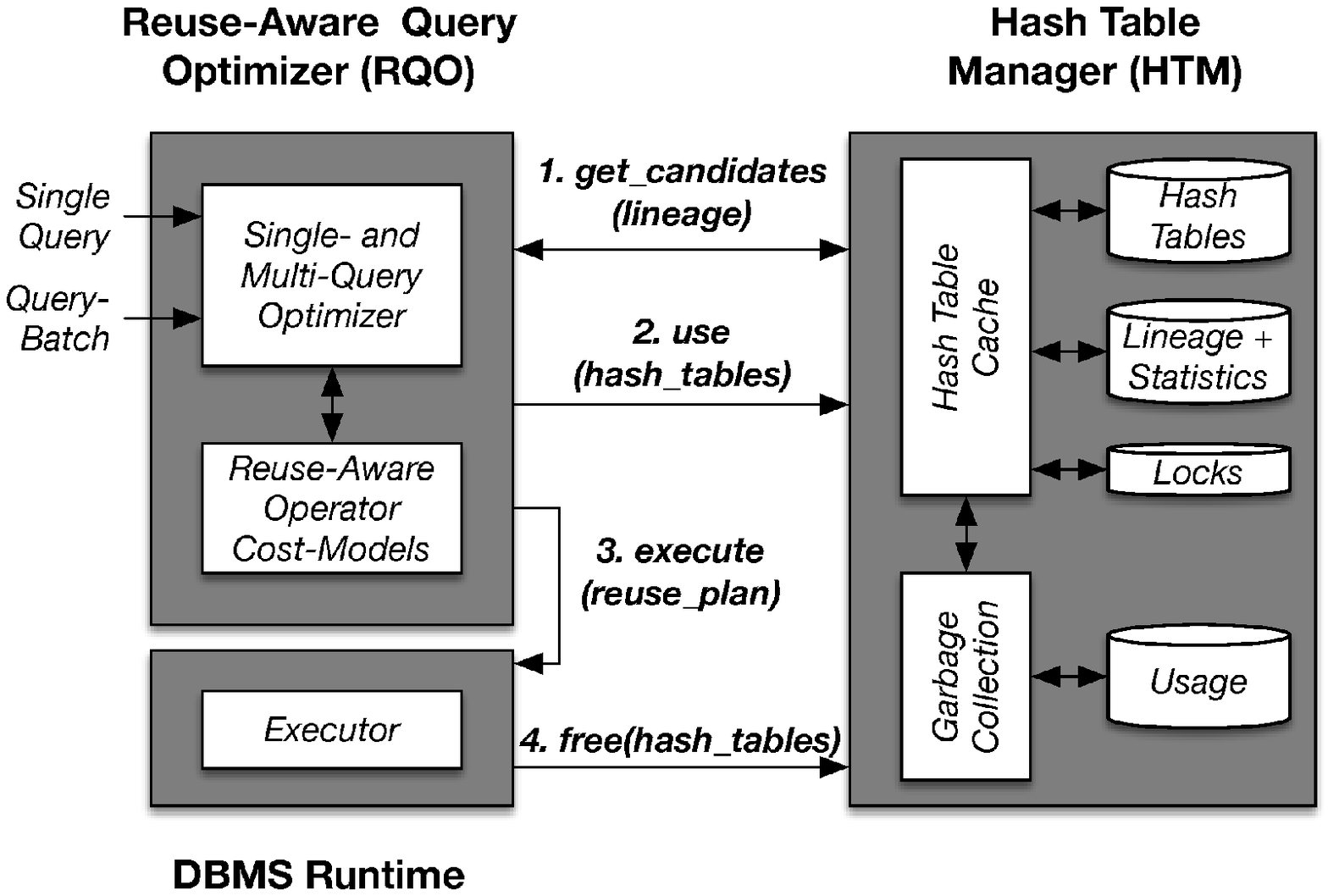}
\caption{Additional \system{} Components}
\vspace{-3.5ex}
\label{fig:arch}
\end{figure}

The main goal of \system{} is to leverage internal hash tables for reuse that are materialized during query execution. 
To achieve this, in \system{} we add the following components to a classical DBMS architecture (see Figure \ref{fig:arch}):
(1) a \emph{Reuse-aware Query Optimizer} (RQO) that replaces the traditional (non-reuse-aware) optimizer, and 
(2) a \emph{Hash Table Manager} (HTM) that consists of a cache of hash tables and a garbage collector. 
In the following, we discuss each component individually and then present an example to illustrate the main ideas of \system{}.

\subsection{Reuse-Aware Query Optimizer}

The Reuse-Aware Query Optimizer (RQO) offers two interfaces for compiling and optimizing queries: a query-at-a-time interface for single-query reuse and a query-batch interface to support multi-query reuse. \\

\vspace{-1.5ex}
\textbf{Query-at-a-time Interface:} This interface accepts a single query, optimizes this query, and returns a reuse-aware execution plan.

The main goal of the reuse-aware optimizer is to decide which hash tables in the cache to reuse such that the overall query execution time is minimized. 
In order to select the best reuse-aware execution plan, the reuse-aware \system{} optimizer enumerates different join orders and decides for each plan which is the best reuse case based on the hash tables in the cache. 
Different from a traditional query optimizer, our reuse-aware optimizer therefore implements two important extensions:
(1) In order to decide which hash table to reuse, the optimizer leverages so called \emph{reuse-aware cost models}.
Different from normal cost models, reuse-aware cost models additionally take statistics of a candidate hash table into account in order to estimate the execution costs for the different reuse cases discussed before.
(2) The reuse-aware optimizer implements benefit-oriented optimizations.
The main intuition is that a plan is preferred over another if it creates hash tables that promise more benefits for future reuse even if the initial execution is a ``little'' more expensive.

Furthermore, the \system{} optimizer supports four different cases for reuse-aware operators: \emph{exact}-, \emph{subsuming}-, \emph{partial}-, and \emph{overlapping}-reuse.
This is different from the existing approaches in \cite{monetdb2009,cod2001,vvreuse_13}, which only support the \emph{exact}-reuse, and the \emph{subsuming}-reuse cases.
The \emph{exact} case enables a join or aggregation operator to reuse a cached hash table which contains exactly the tuples required by the query.
In that case complete sub-plans might be eliminated (e.g., the one which build the hash table of a hash-join).
Compared to the case before, the \emph{subsumption} case is possible when the reused hash table contains more tuples than needed. This might lead to false-positives, which need to be post-filtered by an additional selection.
The \emph{overlapping} and the \emph{partial} case are different.
Both cases allow the reuse of a hash table where some tuples are  ``missing''. These tuples thus need to be added during query execution.
To support all these different reuse cases the optimizer also applies different rewrites rules during optimization.

Figure \ref{fig:arch} shows how the reuse-aware optimizer is integrated into the \system{} architecture.
First, the optimizer enumerates different join orders and retrieves candidate hash tables for reuse.
Once the "optimal" reuse-aware execution-plan is found, the optimizer sends the information of which hash tables will be reused to the HTM for book-keeping as shown in step 2 (Figure \ref{fig:arch}). 
Finally, the optimizer sends the reuse-aware execution plan to the executor as shown in step 3 (Figure \ref{fig:arch}).
Once the plan execution is finished, the DBMS runtime informs the HTM to release all used reused hash tables as shown in step 4 (Figure \ref{fig:arch}), which make them available for garbage collection for instance.
Details about the query-at-a-time interface are described in Section  \ref{sec:single}. \\

\vspace{-1.5ex}
\textbf{Query-Batch Interface:} The query-batch interface is different from the query-at-a-time interface since it accepts multiple queries submitted as a batch.
Different from the query-at-a-time interface, subsets of queries submitted in the same batch can share the same execution plan; called \emph{reuse-aware shared plan} in the sequel.
Different from the approach presented in \cite{sharedb2014}, the main contribution of \system{} is that it integrates the before-mentioned reuse techniques into shared plans.
In order to find the best reuse-aware shared plan, we developed a novel reuse-aware multi-query optimizer that merges individual reuse-aware plans using a dynamic programming based approach.
Details about the query-batch interface are described in Section  \ref{sec:multi}.

\subsection{Hash Table Manager}

The two components of the Hash Table Manager (HTM) are the hash table cache and the garbage collector.\\

\vspace{-1.5ex}
\textbf{Hash Table Cache:}
The hash table cache manages hash tables for reuse; it stores pointers to cached hash tables, as well as lineage information about how each one of them was created. 
It also stores statistics to enable the cost-based hash table selection by the optimizer.
For our initial prototype, we allow only one query to reuse a hash-table in the cache at a time (except for the query-batch interface).
However, for future work, we plan to look into sharing the same hash table between concurrent queries. \\

\vspace{-1.5ex}
\textbf{Garbage Collector:}
The main goal of the garbage collector is to decide which hash tables to evict as necessary.
The garbage collector is triggered by the cache if no more memory is available to admit new hash tables.
Therefore, the garbage collector maintains usage information and implements eviction strategies to decide which hash tables can be removed. 

\subsection{Reuse Example}

\begin{figure}
\centering
\includegraphics[width=0.45\textwidth]{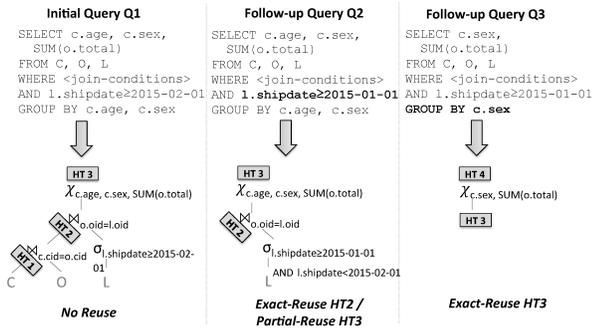}
\caption{Reuse Example}
\vspace{-3.5ex}
\label{fig:example}
\end{figure}

Figure \ref{fig:example} illustrates a reuse example for a sequence of three queries from a data exploration session.
The initial query $Q1$ executes an aggregation over a 3-way join of the tables \texttt{Customer}, \texttt{Orders}, and \texttt{Lineitem} for all lineitems shipped after \texttt{2015-02-01}.
For this query no reuse is possible (since it is the initial one).
However it materializes all three hash tables HT1-HT3 in the cache of \system{}.

The follow-up query $Q2$ then executes a query that differs from $Q1$ only in the the filter predicate; i.e., it selects lineitems that shipped after \texttt{2015-01-01}.
In order to execute $Q2$, \system{} can reuse hash table $HT2$ (exact-reuse) and thus avoids to recompute the join of \texttt{Customer} and \texttt{Orders}. 
Moreover, the hash table $HT3$ produced by the aggregation of $Q1$ can also be reused to compute the aggregation operator in $Q2$ well. However, since $HT3$ does not aggregate all required lineitems (due to partial-reuse), the base table \texttt{Lineitem} needs to be re-scanned for the ``missing'' tuples between \texttt{2015-01-01} and \texttt{2015-02-01}. These tuples are added to $HT3$ by the reuse-aware plan of $Q2$.

Finally, the last query $Q3$ is similar to $Q2$.
The only difference is that it removes the group-by attribute \texttt{c.age}.
For executing $Q3$, \system{} can directly reuse the hash table $HT3$ (exact-reuse).
However, due to the removed group-by attribute a post-aggregation operator needs to be added.

\vspace{-1.5ex}
\section{Single-query Reuse}
\label{sec:single}

In this section, we describe how to find the best reuse-aware execution plan for the query-at-a time interface.
As discussed before, finding the best reuse-aware plan is implemented by the optimizer of \system{}.
Therefore, we first give an overview of how the plan enumeration procedure of our optimizer works and then discuss the cost models of our reuse-aware hash-join and hash-aggregate operator.
Afterwards, we present the details on how the matching and rewriting procedures in \system{} work to enable the different reuse-cases (exact, subsuming, partial, and overlapping).
Finally, we discuss some benefit-oriented optimizations that increase the effect of reuse by spending initially a little more execution cost to create "better" hash tables.

\subsection{Reuse-aware Plan Enumeration}
\label{sec:single:pe}

The plan enumeration algorithm in~\system{} can be applied to complex nested SQL queries. 
In order to simplify the presentation, we first show the basic procedure that only enumerates different join plans for a given SPJ (select-project-join) query .
More complex queries including aggregations and nesting are discussed at the end of this section. \\

\vspace{-1.5ex}
\textbf{Basic Procedure:} Algorithm \ref{alg:planenum} shows the basic recursive procedure for enumerating different reuse-aware plans for SPJ queries based on a top-down partitioning search strategy.
Compared to existing top-down partitioning search strategies such as \cite{topdown2007}, Algorithm \ref{alg:planenum} additionally implements the following ideas to support reuse-aware plans:
(1) when partitioning the join graph into a left and right partition of $G$, the algorithm enumerates the different candidate hash tables (including a new empty hash table) for the right partition $G_r$ and the left partition $G_l$ that can be reused for building the hash table of a hash join (line 8 to 16 and 19 to 27).
(2) Another difference from existing top down enumeration algorithms is to rewrite the respective sub-plan that would reuse a given candidate hash table (line 9 and 20). 
This rewrite possibly eliminates the complete sub-plan (i.e., in the best case $G'_r$ (line 9) and $G'_l$ (line 20) might become an identity operation over the reused hash table if an exact-reuse is possible). 
We discuss details of the rewrite procedure for all four different reuse cases (exact-, subsuming-, overlapping-, and partial-reuse) later in this section.
(3) The last difference is that the cost estimation (line 13 and 24) uses the cost models for the reuse-aware join and aggregation operator to estimate the runtime costs when reusing a given candidate hash table.
We also discuss the details of these cost models later in this section.

\begin{algorithm}
\scriptsize
\SetAlgoLined
\SetKwInOut{Input}{Input}\SetKwInOut{Output}{Output}
\SetKwFunction{optimize}{getBestReusePlan}
\SetKwFunction{genPlan}{generatePlans}
\SetKwFunction{buildTree}{buildTree}
\SetKwFunction{getGraph}{getJoinGraph}
\SetKwFunction{createTree}{createTree}
\SetKwFunction{rewriteTree}{rewriteTree}
\SetKwFunction{checkHash}{getCandHTs}
\SetKwFunction{createHashSpec}{createHTSpec}
\SetKwFunction{newHash}{createNewHT}
\SetKwFunction{cost}{cost}
\SetKwData{Left}{left}
\Input{Join Graph $G$ of SPJ Query $Q$}
\Output{Reuse-Aware Execution Plan $P$}
\BlankLine
\textbf{Algorithm} {\optimize{$G$}}\\
{
\eIf{$bestPlans[G] \neq NULL$}{
\KwRet{$bestPlans[G]$}\;
}
{
\ForEach{$partition$ $(G_l,G_r)$ in $G$}{
$candHTs \leftarrow \checkHash(G_r)$\;
$candHTs \leftarrow candHTs \cup \newHash(G_r)$\;
\BlankLine
\ForEach{$candHT \in candHTs$}{
$G_r' \leftarrow \rewriteTree(G_r,candHT)$\;
$P_l \leftarrow \optimize(G_l)$\; 
$P_r' \leftarrow \optimize(G_r')$\;
$curTree \leftarrow \createTree(P_l, P_r', candHT)$\;
\If{$\cost(curTree) \leq \cost(bestPlans[G])$}{
$bestPlans[G] = curTree$\;
}
}
\BlankLine
$candHTs \leftarrow \checkHash(G_l)$\;
$candHTs \leftarrow candHTs \cup \newHash(G_l)$\;
\BlankLine
\ForEach{$candHT \in candHTs$}{
$G_l' \leftarrow \rewriteTree(G_l,candHT)$\;
$P_l' \leftarrow \optimize(G_l')$\; 
$P_r \leftarrow \optimize(G_r)$\;
$curTree \leftarrow \createTree(P_r, P_l', candHT)$\;
\If{$\cost(curTree) \leq \cost(bestPlans[G])$}{
$bestPlans[G] = curTree$\;
}
}
\BlankLine 
}
}
}
\caption{Plan Enumeration in \system{}}
\label{alg:planenum}
\end{algorithm}

\textbf{Complex Queries:}
The general idea to support more complex queries is similar to  exiting optimizers. 
First, nested queries are unnested using joins if possible.
Second, if unnesting is not possible for the complete query, the enumeration procedure shown before is applied to each query block individually. In \system{}, query blocks can be in the form of either SPJ (select-project-join) or SPJA (select-project-join-aggregation) queries.

In order to extend Algorithm \ref{alg:planenum} for SPJA queries, we only need to iterate over all candidate hash tables as well as an empty (new) hash table for the additional aggregation operator in the SPJA block. 
After selecting the most optimal hash table for reuse, we then need to apply the rewrite rules for the aggregation operator once the procedure shown in Algorithm \ref{alg:planenum} returns its result.

\subsection{Reuse-Aware Operators and Cost Models}
\label{sec:single:rop}

In the following, we discuss the reuse-aware operators (join and aggregation) and the cost models to select hash tables for reuse.

\vspace{-0.75ex}
\subsubsection{Reuse-Aware Hash-Join}
\label{sec:single:rhj}

A reuse-aware hash-join (RHJ) works similarly to a traditional hash-join; i.e., the join first builds a hash table from one of its inputs and then probes into the hash table using each tuple of the other input.
However, an RHJ has two major differences:
(1) in the build phase, the RHJ operator might need to add the ``missing'' tuples into the reused hash table (for overlapping- and partial-reuse), and  
(2) in the probe phase, the RHJ operator might need to post-filter false-positives (for overlapping- and subsuming-reuse); i.e., tuples that are stored in a reused hash table but not required to execute the current join operator.
Running the join without post-filtering would produce false-positives during the probing phase.

For each candidate hash table $HT$ that can be reused to compute a given join, the optimizer in \system{} needs to estimate the total runtime costs. 
In the following, we explain the details of our cost model.\\

\vspace{-1.5ex}
\textbf{Cost Model:} The main components that determine the cost of an RHJ are the resize cost $c_{resize}$, the build cost $c_{build}$ and the probe cost $c_{probe}$.  

\vspace{-1.5ex}
\begin{scriptsize}
\begin{equation*}
c_{RHJ} = c_{resize}(HT) + c_{build}(HT) + c_{probe}(HT)
\end{equation*}
\end{scriptsize}
\vspace{-1.5ex}

Our cost model for the RHJ explicitly considers the cost for resizing the hash table, $c_{resize}$.
In order to minimize the cost of resizing in \system{}, we use a hash table that implements extendible hashing using linked lists for collision handling.
Thus, instead of re-hashing all entries, only the bucket array needs to get resized and entries can be assigned to the new buckets lazily.

The costs for building and probing of an RHJ are different from a traditional hash-join and depend additionally on two parameters: 
(1) the contribution-ratio $contr$ and 
(2) the overhead-ratio $overh$ of a candidate hash table $HT$.
The first parameter, the contribution-ratio $contr$, defines how much of the data in the candidate hash table $HT$ already contributes to the operator if that operator reuses this hash-table; i.e., this data does not need to be added to the hash table anymore and makes the build phase faster.
For example, if $contr(HT)=0.5$ then only $50\%$ of missing tuples need to be added to the hash table $HT$ during the build phase.
The second parameter, the overhead-ratio $overh$, defines how much unnecessary data is stored in the hash table; i.e., this data contributes to the total memory footprint of the hash table and makes the building and probing phases slower since the hash table might spill out of the CPU caches.
The overhead-ratio also determines the additional cost needed to post-filter false positives.
For example, if $overh(HT)=0.7$ then $70\%$ of tuples in the hash table are not required by the RHJ.
In the sequel, we discuss how to use both parameters ($contr$ and $overh$) to model all of the four reuse-cases (exact, subsuming, partial, and overlapping).

In the following equations, we show how \system{} estimates the costs of the build phase $c_{build}(HT)$ and the probe phase $c_{probe}(HT)$ of an RHJ using these two parameters.

\vspace{-1.5ex}
\begin{scriptsize}
\begin{equation*}
\begin{split}
c_{build}(HT) &=  \underbrace{|Builder| \cdot (1-contr(HT))}_\text{\#tuples to insert} \cdot \underbrace{ c_i(htSize,tWidth)}_\text{cost of a single insert} \\
c_{probe}(HT) &=  \underbrace{|Prober|}_\text{\#tuples to probe}\cdot \underbrace{c_r(htSize,tWidth)}_\text{cost of a single lookup}
\end{split}
\end{equation*}
\end{scriptsize}
\vspace{-1.5ex}

The build cost $c_{build}(HT)$ is determined by the number of missing tuples that need to be inserted times the cost of a single insertion $c_i$ into the resized hash table.
The probe cost $c_{probe}(HT)$ is determined by the number of tuples that need to probe into the hash table times the lookup cost $c_l$ for a single probe into the hash table.

The cost of a single insert/lookup $c_i$ and $c_l$ depend on two parameters: 
(1) the memory footprint of the resized hash table $htSize$ (shown in the following equation) and
(2) the width of a tuple $tWidth$ stored in the cached hash table $HT$.
While the memory footprint $htSize$ determines if a hash table fits into the CPU caches or not and thus influences the insert/lookup costs, the tuple width $tWidth$ determines the number of I/O operations required to transfer a tuple between main memory and CPU caches.
Since the hash table might contain more attributes than needed by the query (e.g., for post-filtering), the tuple-width $tWidth$ might actually be bigger as for a hash table that we create individually for this query.

\begin{figure*}
  \begin{minipage}[b]{.25\textwidth}
    \centering
    \includegraphics[width=\textwidth,trim=5mm 0mm 4mm 0mm]{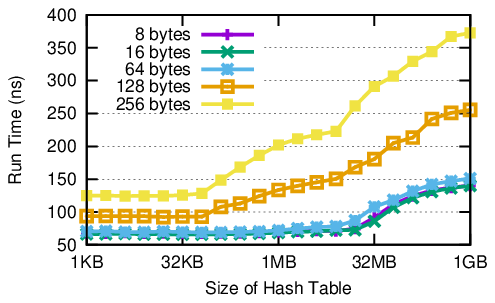}
    \vspace{-3.5ex}\subcaption{Cost of a single insert}
    \label{fig:single_insert}
  \end{minipage}
  \hfill
  \begin{minipage}[b]{.25\textwidth}
    \centering
    \includegraphics[width=\textwidth,trim=5mm 0mm 4mm 0mm]{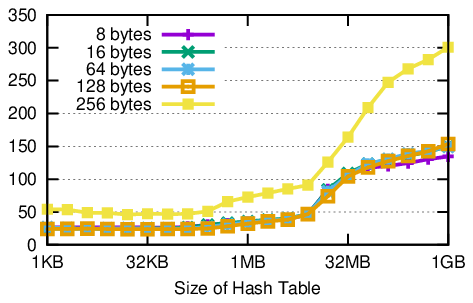}
    \vspace{-3.5ex}\subcaption{Cost of a single probe}
    \label{fig:single_probe}
  \end{minipage}
  \hfill
  \begin{minipage}[b]{.25\textwidth}
    \centering
    \includegraphics[width=\textwidth,trim=5mm 0mm 4mm 0mm]{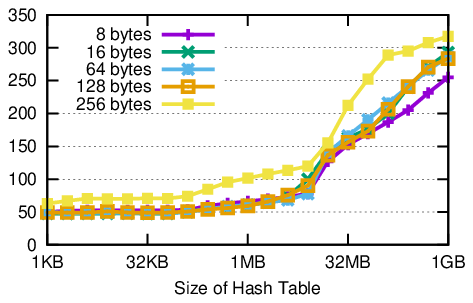}
    \vspace{-3.5ex}\subcaption{Cost of a single update}
    \label{fig:single_update}
  \end{minipage}
  \hfill
\\
  \vspace{-1.5ex}
  \caption{Reuse-aware Cost Parameters}
  \vspace{-3.5ex}
  \label{fig:cost_model_param}
\end{figure*}

Moreover, for estimating the build cost $c_{build}$ and the probe cost $c_{probe}$, we need to be able to estimate the cost of a single insert/lookup ($c_i$ and $c_l$).
However, these costs depend on the specific hash table implementations and other hardware-dependent parameters; e.g., how prefetching into CPU caches is implemented.
Therefore, these costs need to be determined by a set of micro benchmarks which calibrate the cost model for varying hash table sizes and tuple widths.
We implemented such a micro benchmark using C++ and used GCC 4.9.2 as compiler.
Figure \ref{fig:single_insert} and Figure \ref{fig:single_probe} show the results of our micro benchmark for the hash table implementation used in \system{} on a machine with an Intel Xeon E5-2660 v2 processor using $128$GB RAM running Ubuntu Server 14.04.1.
The cache sizes of the processor are: $32$KB data + $32$KB instruction L1 cache (private), $256$KB L2 cache (private) and $25$MB L3 cache (shared).

For both, the insert and lookup operations, we can clearly see the effect of different hash table sizes ($1KB$ to $1GB$) and cache boundaries on the insertion/lookup costs. 
The effect of the tuple-width ($8B$ to $256B$) is also visible but needs some more explanation.
For insertion, the cost does not change as long as a tuple fits into one cache line, which is $64$B in our processor.
Once the tuple-width exceeds the cache line size, the cost increases as shown for $128$B and $256$B in Figure \ref{fig:single_insert}.
For lookup, the behavior is slightly different: due to the prefetching of one cache line by the CPU, the cost to lookup one tuple increases only when the tuple-width exceeds $128$B.

\vspace{3.5ex}
\subsubsection{Reuse-Aware Hash-Aggregation}
\label{sec:single:rha}

Similar to the reuse-aware hash-join (RHJ), the reuse-aware hash-aggregate (RHA) can reuse an existing cached hash table.
Similar as for the RHJ, the RHA might also need to add ``missing'' tuples (for overlapping- and partial-reuse) and post-filter tuples (for overlapping- and subsuming-reuse).
In the following, we discuss the cost model, that estimates the runtime cost of an RHA for a given hash table.
\\

\vspace{-1.5ex}
\textbf{Cost Model:}
For a given candidate hash table, the optimizer estimates the total runtime costs of a reuse-aware hash aggregate as shown by the following equation: 

\vspace{-1.5ex}
\begin{scriptsize}
\begin{equation*}
c_{RHA} = c_{resize}(HT) + c_{insert}(HT) + c_{update}(HT)
\end{equation*}
\end{scriptsize}
\vspace{-1.5ex}

The cost of an RHA consists of three components: 
(1) the resize cost $c_{resize}$,
(2) the cost $c_{insert}$ to insert the initial tuple for each distinct  missing group-by key, and 
(3) the update costs $c_{update}$ of the aggregated value for the other input tuples.
For example, assume an RHA has $100$ missing input tuples with $10$ distinct missing group-by keys.
In that case, the RHA needs to pay $10$ times the insert cost and $90$ times the update cost to reuse the given hash table.
Similar to the RHJ, the contribution-ratio $contr$ and the overhead-ratio $overh$ have an influence on the insert/update costs.

In the following, we take a closer look into defining different cost components for a given candidate hash table $HT$. 
The cost component $c_{resize}$ represents the cost to resize the hash table for the distinct missing group-by keys. 
Again, these costs are dependent on the implementation details of the hash table.

For RHA, we use the same hash table implementation as for the RHJ operator (i.e., we use the same cost estimates for resizing).
The idea to estimate the other two cost components is that the insertion cost only need to be paid for the input tuples that represent the missing distinct group-by keys (i.e., the first tuple that is inserted for a missing key). All other tuples need to pay only the update cost. 
The functions to estimate the insertion cost $c_{insert}$ and the update cost $c_{update}$ are shown in the following equations.

\vspace{-1.5ex}
\begin{scriptsize}
\begin{equation*}
\begin{split}
c_{insert}(HT) &=  \underbrace{|distinct(Input.key)| \cdot (1-contr(HT))}_\text{\#tuples to insert} \cdot \\ 
& \underbrace{ c_i(htSize,tWidth)}_\text{cost of a single insert}
\end{split}
\end{equation*}
\end{scriptsize}

\begin{scriptsize}
\begin{equation*}
\begin{split}
c_{update}(HT) &=  \underbrace{(|Input|-|distinct(Input.key)|) \cdot (1-contr(HT))}_\text{\#tuples to update} \cdot \\
&\underbrace{c_u(htSize,tWidth)}_\text{cost of a single update}
\end{split}
\end{equation*}
\end{scriptsize}
\vspace{-1.5ex}

The equations above need an estimate for the insert/update cost ($c_i$, $c_u$) for a input single tuple.
Same as for the RHJ, these costs depend on the total size of the hash table  $htSize$ and the width of the tuples $tWidth$ stored in the candidate hash table $HT$ as well as the other hardware-dependent factors (i.e., the CPU cache sizes).
Different from the RHJ, however, the size of the hash table $htSize$ only depends on the number of distinct keys in the input since only one aggregated tuple per distinct key need to be stored.

In order to calibrate our costs for our hash-table implementation and the underlying hardware, we again executed a set of micro-benchmarks.
The insertion costs $c_i$ are the same as the ones shown for the RHJ operator in Section \ref{sec:single:rhj}.
Figure \ref{fig:single_update} shows the results for the update costs $c_u$ (single-threaded) for the same setup (CPU, hash table implementation and compiler) used for the RHJ operator.

\subsection{Matching and Rewriting}
\label{sec:single:rw}

The goal of the matching procedure is to find those candidate hash tables in the hash table cache which qualify for reuse for a given operator $r$ in the enumerated plan.
As discussed before, in \system{} we support reuse for hash-join and hash-aggregations only.
We use the following notation: $C$ (cached plan) represents the plan that produced a \emph{cached} hash table and $R$ (requesting plan) represents the plan rooted by $r$.
For a hash-join $r$, $R$ represents the sub-plan below the join input that builds the hash-table (including the join itself but excluding the probing branch).
For a hash-aggregate $r$, the sub-plan $R$ represents the operator tree below the aggregation operator including the aggregation itself.

\begin{figure}
\centering
\includegraphics[width=0.30\textwidth]{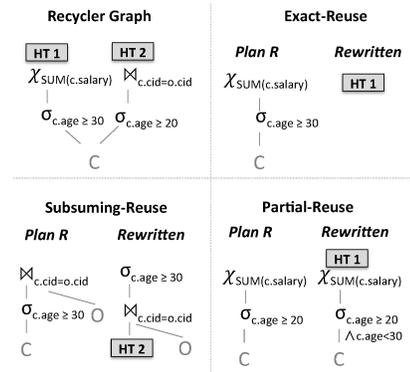}
\vspace{1.5ex}
\caption{Match and Rewrite}
\vspace{-3.5ex}
\label{fig:recycler}
\end{figure}

For finding a matching hash table that can be reused for $r$, our hash table manager stores lineage information in a similar way as described in \cite{vvreuse_13} using a so called the recycle graph $G_C$. 
The graph $G_C$ merges the lineage of all hash tables in one graph $G_c$.
Figure \ref{fig:recycler} shows an example of a recycler graph that resulted from merging two query plans; the first query that produces $HT1$ contains an aggregation operator over the \texttt{Customer} table and the second query plan contains a join operator that builds the hash table $HT2$ over the \texttt{Customer} table on the join key \texttt{cid}.
Different from the recycler graph in \cite{vvreuse_13}, for each node $n_c \in G_C$ we additionally store if it refers to a cached hash table or not (since not all operators materialize a hash table).

The problem of matching is now to find a subtree in $G_C$ that matches a given plan $R$ rooted by the requesting operator $r$.
Matching $R$ to the graph $G_C$ builds on the notion of bisimilarity; i.e., all nodes $n_c$ in the graph $G_C$ are compared to the root node $r$ of $R$.
In our case, we can prune the search space and only need to compare $r$ to those nodes $n_c$ that actually refer to a cached hash-table (i.e., all operators that represent joins and aggregates in $G_c$).

The matching procedure used in \cite{vvreuse_13} to check if a node $n_c$  exactly matches $r$ is: (i) $n_c$ and $r$ represent the same type of operation (e.g., both are selections); (ii) the parameters of the two nodes are equal (e.g., they evaluate the same selection predicate); (iii) they have the same number of children and there is an exact match for all children of $n_c$ and $r$. 
This procedure, however, tests only for the exact-reuse case.
\cite{vvreuse_13} also discusses an extension to test for subsuming-reuse and not only for exact-reuse. 
In the following, we explain, how we extend the above matching procedure to check for the all four reuse-cases in the given order and discuss which rewrites need to be applied in each reuse case.
\\

\vspace{-1.5ex}
\textbf{Exact-reuse:} The matching procedure detects an exact-reuse of the hash-table referred by $n_c$, if the predicates of all selection operators in $R$ match exactly one predicate in the query tree rooted by $n_c$; i.e., for each predicate $r_i$ in $R$ there exists a predicate $c_i$ in the query tree rooted by $n_c$ for which $r_i = c_j$ holds.
In that case $R$ can be replaced by the operator over the cached hash table that is referenced by $n_c$.
Figure \ref{fig:recycler} (top right) shows an example where the selection predicate $\sigma_{c.age \geq 30}$ of $R$ matches the sub-plan rooted by the aggregation operator in the recycler graph $G_c$. 
The rewrite rule thus replaces the original plan $R$ directly by $HT1$.

In case that the root node $r$ of $R$ represents a hash-aggregation, we additionally allow that the group-by attributes $G_r$ of $r$ are only a subset of the group-by attributes $G_c$ of $n_c$ if all aggregation functions are additive (e.g., sum, count, min, max).
In that case, $R$ can be replaced by a sub-plan that consists of the aggregation operator $r$ over the reused hash-table.
Figure \ref{fig:example} (right hand side) already has shown an example for this reuse case.\\

\vspace{-1.5ex}
\textbf{Subsuming-reuse:} We test if there exists a plan $C$ in $G_c$ where $r_i \subseteq c_j$ holds for all selection predicates $r_i$ in the plan rooted by $r$.
In that case $R$ can be replaced by a selection operator $\sigma_{post}$ over the reused hash table.
The predicate $post$ represents the conjunction of all predicated $r_i$ of $R$.
If the hash table does not contain the attributes needed to test $post$, it does not qualify for reuse.
Figure \ref{fig:recycler} (bottom left) shows this case. Since the hash table $HT2$ contains all customers for $age \geq 20$ and the join $r$ only requires only customer with $age \geq 30$, all false positives must be post filtered after probing using the filter predicate $\sigma_{age \geq 30}$.
\\

\vspace{-1.5ex}
\textbf{Partial-reuse:} In order to support this reuse case, we test if there exists a plan $C$ in $G_c$ where $c_j \subseteq r_i$ holds for all selection predicates $r_i$ in the plan rooted by $r$.
This means, that the reused hash table does not contain all necessary tuples.
To the end, $R$ is rewritten to a plan $R'$ which adds the missing tuples from the base tables to the reused hash table by replacing $r_i$ in $R$ by $r_i \wedge !c_j$.
Figure \ref{fig:recycler} (bottom right) shows an example where $HT2$ can be partially reused; i.e., all customers  $20 \geq age < 30$ must be added to $HT2$ from the base table \texttt{Customer}.
\\

\vspace{-1.5ex}
\textbf{Overlapping-reuse:} We test if there exists a plan $C$ in $G_c$ where $c_j \wedge r_i \ne \emptyset$ holds for all selection predicates $r_i$ in the plan rooted by $r$.
For rewrite, we apply both rewrites that we have discussed for the partial-reuse and the subsuming-reuse case before.

\subsection{Benefit-oriented Optimizations}
\label{sec:single:benefit}

\system{} additionally implements the following benefit-oriented optimizations.
The main intuition behind these optimizations is that one plan $A$ is preferred over another plan $B$ if the plan $A$ creates hash tables that promise higher benefits for future reuse even if the plan execution is a ``little'' more expensive then for the optimal plan.\\

\vspace{-1.5ex}
\textbf{Aggregate Rewrite:} AVG is always rewritten to SUM and COUNT to support the partial- and overlapping-reuse at the cost of initially creating a slightly bigger hash-table.  \\

\vspace{-1.5ex}
\textbf{Additional Attributes:} For join hash tables, all attributes used in selection operations in the sub-plan of the input which build the hash-table are added to the cache as well to enable post-filtering of false positives to increase the reuse potential of a hash table. \\

\vspace{-1.5ex}
\textbf{Join Order:} Typically hash tables are always built over the smaller join input. However, if the hash table is reused in future it might be also beneficial to build the hash table over the bigger input. 
We therefore integrated a simple heuristic approach into our optimizer that is similar to the techniques presented in \cite{vvreuse_13} to determine which intermediate will provide more benefit for future queries based on the history of queries executed.\\
\vspace{-1.5ex}
\section{Multi-query Reuse}
\label{sec:multi}


In this section, we describe the techniques in \system{} that enable shared plans to reuse cached hash tables.
We call these plans \emph{reuse-aware shared plans}.
In the following, we first discuss the details of reuse-aware shared plans.
Afterwards, we present how we extend our optimizer in \system{} to find an optimal reuse-aware shared plan for a given query-batch and a set of cached hash tables.

\subsection{Reuse-Aware Shared Plans}

\begin{figure}
\centering
\includegraphics[width=0.5\textwidth]{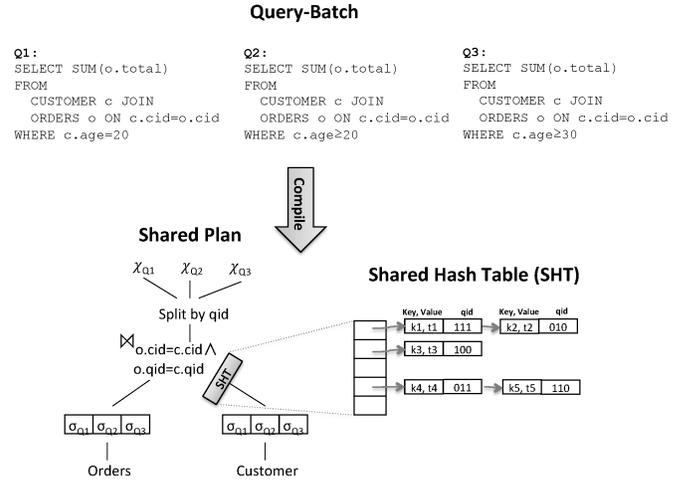}
\caption{Shared Plans using a Data-Query Model}
\vspace{-4.5ex}
\label{fig:shared}
\end{figure}

The basic idea of shared plans is shown in Figure \ref{fig:shared}.
Instead of compiling each query into a separate plan, multiple queries are compiled into one shared plan that reuses hash tables.
The idea of shared plans has been presented in \cite{sharedb2014} already.
In \system{}, we extend shared plans to allow them to reuse cached hash tables.
In the following, we first reiterate over the idea of shared plans and then discuss the relevant modifications for our reuse-aware operators to work correctly in shared plans.

Different from a normal plan, in a shared plan individual operators execute the logic of multiple queries.
The most common shared operator is the shar\-ed scan operator that evaluates the filter predicates of multiple queries in one scan.
In order to keep track of which tuples qualify for which query, shared operators in \cite{sharedb2014} use a Data-Query Model where each tuple is tagged by the IDs of those queries it qualifies for.
For example, if a tuple produced by a shared scan satisfies the predicates of query $Q_1$ and $Q_3$ but not of query $Q_2$, this tuple will be tagged using $Q_1$ and $Q_3$ (or $101$ if a bitlist is used to represent query IDs).
Moreover, other operators such as joins and aggregation operators can be shared as well.
Figure \ref{fig:shared} shows an example of a shared plan where the selection operators and the hash-join are shared by three queries ($Q_1$ to $Q_3$) while the aggregation is not shared (i.e., there exists one separate operator for each query).
For the hash-join, we see that tuples tagged with query IDs (qids) are stored in its hash table. 
The query IDs are used in the probing phase to produce the output of the join.

In the following, we describe our extensions for the reuse-aware hash-join and hash-aggregate such that they can execute multiple queries at a time.\\

\vspace{-1.5ex}
\textbf{Shared Reuse-Aware Hash-Joins:}
In general, the shar\-ed reuse-aware hash-join (SRHJ) operator works similarly to the non-shared reuse-aware hash-join (RHJ) presented in Section \ref{sec:single:rhj}:
Instead of recomputing the hash table in the build phase from scratch, a cached hash table is reused to avoid re-computation.

However, there are some important differences between an SRHJ that has to support query-batches and a non-shared RHJ that only supports a single query.
First, the SRHJ can only reuse hash tables that include query IDs for tagging (as shown in Figure \ref{fig:shared}).
A hash table that does not include query IDs can not be reused for a shared operator.
Second, before the SRHJ operator starts to execute it has to re-tag all tuples stored in the reused hash table using the predicates of current query-batch.
Otherwise, if it does not re-tag all tuples in the reused hash table, these tuples will be tagged with obsolete query IDs from a previous (non-active) query-batch, which might lead to wrong query results if query IDs are recycled. 
To that end, re-tagging represents an overhead that has to be considered in the cost model of an SRHJ.\\

\vspace{-1.5ex}
\textbf{Shared Reuse-Aware Hash-Aggregates:}
Shared aggregates are different from normal aggregation operators since they split the execution into two phases: 
a first phase that groups the input tuples by keys and a subsequent aggregation phase.
While the grouping phase is shared for all queries, the subsequent aggregation phase is carried out for each query separately (i.e., the output of the grouping phase is split based on query IDs).
In this paper, we focus on shared hash-aggregates that store the output of the grouping phase in a hash table before applying the aggregation functions on the individual tuples stored in the hash-table.

The goal of a shared reuse-aware hash-aggregate (SRHA) is to reuse hash tables to avoid the re-computation of the grouping phase.
This is very different than reusing hash tables for a non-shared RHA operator since hash tables of an SRHA store individual tuples and not aggregates.
Another difference is that the SRHA operator also needs to re-tag all the tuples stored in the reused hash table (just as for the SRHJ operator) before the operator is executed.
Both these aspects (i.e., storing individual tuples and the need for re-tagging) influence the cost of an SRHA and must be included in the cost model.

Finally, SRHA and the RHA operators also differ in how they select candidate hash tables from the cache.
While an RHA must find hash tables with the same aggregation functions, an SRHA is more flexible since it can recompute any arbitrary aggregate function on the grouped data.
For example, a hash table which was built for an SRHA operator that computes one aggregation function (e.g., SUM) can be reused by another SRHA operator, which computes a different aggregation function (e.g., MIN).

\subsection{Plan Enumeration}

In the following, we discuss the plan enumeration implemented in \system{} to support query-batches.
The goal of the optimizer is to find a set of reuse-aware shared plans $\{S_1, S_2,$ $\dots, S_n\}$ for a given query-batch $\{Q_1,$ $Q_2, \dots, Q_m\}$ with $n \leq m$ that minimizes the total runtime to execute all queries in the given batch by reusing cached hash tables.

\begin{figure}
\centering
\includegraphics[width=0.25\textwidth]{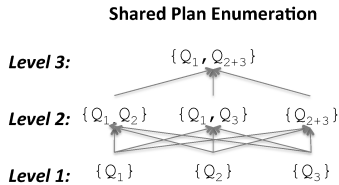}
\vspace{1.5ex}
\caption{Dynamic Programming based Plan Merging}
\vspace{-4.5ex}
\label{fig:sharedplanenum}
\end{figure}

In order to find the optimal set of reuse-aware shared plans $\{S_1, S_2,$ $\dots, S_n\}$, \system{} uses a {\em dynamic programming approach} to merge query plans incrementally into reuse-aware shared plans.  
Figure \ref{fig:sharedplanenum} shows the dynamic programming process for three queries (e.g., such as those in Figure \ref{fig:shared}).
Each node in the dynamic programming graph in \ref{fig:sharedplanenum} represents a so-called merge configuration that describes which queries should be merged together into a shared reuse-aware plan and which should be executed using a separate non-shared reuse-aware plan.
In terms of notation, $\{Q_1,Q_{2+3}\}$ represents a merge configuration, which defines that two separate plans should be generated: one non-shared reuse-aware plan for query $Q_1$ and one shared reuse-aware plan for queries $Q_2$ and $Q_3$.

In \system{}, it depends on two aspects if two queries are merged or not.
First, two queries are merged if the total runtime of the shared plan is less than the sum of executing two individual plans.
Second, not all queries are mergeable.
In order to guarantee a correct plan execution, two queries $Q_1$ and $Q_2$ can only be merged if they have the same join graph. 
Otherwise, these queries cannot be merged and the plans must be kept separate.
If two queries $Q_1$ and $Q_2$ are mergeable, the result of merging in \system{} is a shared reuse-aware plan where (1) all join operations are shared (i.e, SRHJ operators are used for joins) and (2) all aggregation operators that use the same group-by keys are shared (i.e, SRHA operators are used for aggregations).

In order to find the merge configuration that results in the minimal total runtime (i.e., the total sum over all plans), \system{} starts the dynamic programming process with merge configurations of size $1$ (called level $1$).
On level $2$, the optimizer then continues to find the merge configurations for all possible combinations of two queries which has the minimal total runtime by extending the merge configurations from the level below until the process reaches level $m$.
For example, in order to compute the merge configuration on level $3$ in Figure \ref{fig:sharedplanenum}, the dynamic programming process merges query $Q_3$ into the merge configuration $\{Q_1,Q_2\}$ of level $2$ amongst the other possible combinations (e.g., merging $Q_2$ into $\{Q_1,Q_3\}$ or merging $Q_1$ into $\{Q_{2+3}\}$.
In order to merge $Q_3$ with the merge configuration $\{Q_1,Q_2\}$, the dynamic programming process enumerates all three possible merge configurations $\{Q_{1+3},Q_2\}$, $\{Q_{1},Q_{2+3}\}$, and $\{Q_1,Q_2, Q_3\}$ and keeps only the one with the minimal total runtime.
Moreover, in order to avoid to analyze the same merge configuration twice, \system{} memoizes merge configurations and their estimated runtime.

Finally, to estimate the total runtime of a merge configuration, the optimizer computes the optimal reuse-aware (shared) plan associated with each entry of the given merge configuration.
In order to find the best reuse-aware (shared) plan associated with entry in a merge configuration, the optimizer applies a variant of the enumeration process presented in Section \ref{sec:single:pe} that supports query graphs and not only query trees. 
For example, given the merge configuration $\{Q_{1+3},Q_2\}$ the optimizer applies the enumeration procedure to find the best (shared) plan separately for $Q_{1+3}$ and $Q_2$. 
In order to find the best reuse-aware shared plans (e.g., for $Q_{1+3}$), the the plan enumeration in Section \ref{sec:single:pe} uses reuse-aware shared operators (i.e., SRHJ and SRHA) instead of using non-shared reuse-aware operators.

\vspace{-1.5ex}
\section{Garbage Collection}
\label{sec:garbage}

In this section, we provide the details of how garbage collection is implemented in \system{}. 
The main goal of garbage collection is to evict hash tables from the cache that are most likely not to be reused by other queries in future. 
In Section \ref{sec:arch}, we already described that the hash table manager monitors the hash table cache and starts an eviction process, whenever the total memory footprint of the cached hash tables exceeds a threshold (i.e., no more memory is available to store new hash tables). 
To decide on which hash tables to discard is the crucial part of the eviction process. 

Different from the eviction process used in database buffers, the garbage collection in \system{} does not work on the granularity of pages.
Instead it can either work in a coarse-grained mode on the granularity of complete hash tables or in a more fine-grained mode on the granularity of individual hash table entries.
While a coarse-grained mode needs less storage space for book keeping and requires less monitoring overhead than a fine-grained mode, it tends to keep ``old'' entries in a hash table even if other entries in the hash table are only used.
Moreover, evicting individual entries from a hash table in a fine-grained mode requires a scan of individual bucket of the hash table.
Finally, in a fine-grained mode, concurrent access of the eviction process and queries to the same hash table need to be synchronized.

In \system{}, we have implemented this fine-grained mode.
However, initial results showed that this mode results in a high additional load that reduces the efficiency of \system{}.
Therefore, we have decided to integrate only a \emph{least recently used (LRU)} policy that evicts complete hash tables instead of evicting individual entries of hash tables (i.e., garbage collection is working in a coarse-grained mode) in \system{}.
In order to implement the LRU policy, the Garbage Collector of \system{} keeps a \emph{timestamp} of the last access for each hash table in its usage information. 
Based on this timestamp, the eviction process picks the hash table with the oldest timestamp and evicts it from the cache. 
The garbage collection process repeats the eviction until the memory footprint drops below the memory threshold. 
In our experiments we see that this policy is able to efficiently deal with different workloads where queries build on recent results. Moreover, the coarse-grained mode introduces only a minimal overhead for book keeping and for executing the eviction process.

\vspace{-1.5ex}
\section{Experimental Evaluation}
\label{sec:eval}

In this section, we report the results of our experimental evaluation of the techniques presented in this paper. 
The main goal of this evaluation is to:
(1) compare the efficiency of reusing internal hash-tables in \system{} to materialization-based reuse,
(2) present the performance gains for both interfaces: the query-at-a-time and the query-batch interface, 
(3) show the efficiency and the accuracy of our optimizer and the cost models,
(4) analyze the overhead of applying garbage collection in \system{}.

In the following, we explain the details of the experimental setup that are commonly used for all experiments.\\

\vspace{-1.5ex}
\textbf{Workload and Data:} In order to analyze the efficiency of different re-use strategies we are using three different types of analytical workloads with (1) low-, (2) medium-, and (3) high-reuse potential.
Each of the workload consists of $64$ different queries over the TPC-H database schema. 
For the workload with the low-reuse potential, the average overlap of common data read from base tables by two subsequent queries is $1\%$.
This simulates the fact that users often look at different parts of a data set.
For the medium-reuse case, the overlap is $10\%$ and $50\%$ for the high-reuse case.
The idea is that the spatial locality increases in these workloads to simulate users; i.e., in the medium- and high-reuse case users typically explore data in a common region by several queries before changing focus to other parts of the data.

The queries in each workload have the following characteristics: The initial query in each workload is TPC-H query $Q3$ that has a three-way join over the tables \texttt{Lineitem}, \texttt{Orders}, and \texttt{Customer} with an aggregation operator on top.
We used this query since it represents a medium-complex query with three join and one aggregation operator that initially creates three hash tables / intermediate results for potential reuse.
All other queries in our workloads resulted from applying different modifications to simulate different user interactions that are commonly used operations in analytical frontends such as Tableau \cite{tableau2012} or Spotfire \cite{spotfire}. 

The user interactions simulated by different queries are: zooming-in/-out, shifting as well as drill-down and roll-up operations.
While zooming-in/-out and shifting only change the selection predicate of the previous query, drill-down and roll-up add/remove TPC-H tables that use a join and add/remove group-by attributes respectively.
The resulting queries are all SPJ or SPJA queries.

By using \emph{Drill-Down} operations, we introduce new joins into the workload that include \texttt{Part} and \texttt{Supplier} tables to achieve more complex queries (i.e., to form five-way join queries).

Finally, as the main data set in all our experiments, we use a TPC-H database of $SF=10$ with secondary indexes on all selection attributes used in our query workloads.
We do not use other scaling factors since the relative performance gains of \system{} compared to materialization-based reuse is the same.
However, for some of our experiments that contain micro benchmarks, we use synthetic data sets (e.g., to show the effects of our cost models). 
We describe these synthetic data sets further in the corresponding section.
\\

\vspace{-1.5ex}
\textbf{Implementation and Hardware:} 
We implemented our \system{} prototype using C++ and GCC 4.9.2 as the compiler.
For execution in our \system{} prototype, we generate C++ execution plans for all SQL queries following the ideas described in \cite{hyper2011}.
In order to show the pure effects of reuse only, our prototype system implements a single-threaded execution model.
For running all experiments, we used one machine with an Intel Xeon E5-2660 v2 processor and $128$GB RAM running Ubuntu Server 14.04.1.
The cache sizes of the processor are: $32$KB data + $32$KB instruction L1 cache, $256$KB L2 cache and $25$MB L3 cache.

\begin{figure}
  \begin{minipage}[b]{.22\textwidth}
    \centering
    \includegraphics[width=\textwidth, height=0.6\textwidth, trim=5mm 0mm 4mm 0mm]{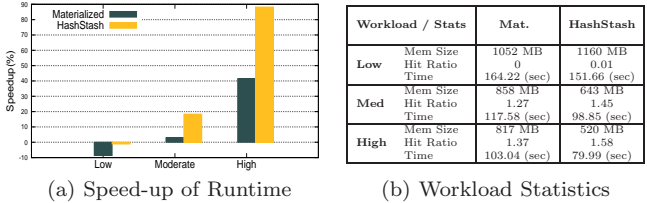}
    \vspace{-3.5ex}
    \subcaption{Speed-up of Runtime}
    \label{fig:exp:runtime}
  \end{minipage}
  \hfill
  \begin{minipage}[b]{.23\textwidth}
    \centering
    \scriptsize
    \resizebox{\columnwidth}{!}{
    \begin{tabular}{|l l |c|c|}
    \hline
    &     & & \\
    \multicolumn{2}{|l|}{\bf Workload / Stats} & {\bf Mat. } & {\bf HashStash} \\
                                &     & & \\\hline 
    \multirow{ 3}{*}{\bf Low}     & Mem Size & $1052$ MB & $1160$ MB \\
                                  & Hit Ratio & $0$ & $0.01$ \\
                                  & Time & 
                                  164.22 (sec) & 
                                  151.66 (sec) \\\hline
    \multirow{ 3}{*}{\bf Med}     & Mem Size & $858$ MB & $643$ MB \\
                                  & Hit Ratio & $1.27$ & $1.45$ \\
                                  & Time & 
                                  117.58 (sec) & 
                                  98.85 (sec)
                                  \\\hline
    \multirow{ 3}{*}{\bf High}    & Mem Size & $817$ MB & $520$ MB \\
                                  & Hit Ratio & $1.37$ & $1.58$ \\
                                  & Time & 
                                  103.04 (sec) & 
                                  79.99 (sec) \\\hline
    \end{tabular}
    }
    \subcaption{Workload Statistics}
    \label{fig:exp:stats}    
  \end{minipage}
  \vspace{0.5ex}
  \caption{Single-Query Reuse}
  \vspace{-4.5ex}
  \label{fig:exp:single}
\end{figure}

\subsection{Exp. 1: Single-Query Reuse}
\label{sec:eval:exp0}

In this experiment, we analyze the benefits of \system{} for the single-query interface using the different workloads mentioned before.
In order to show the efficiency of \system{}, we first executed each workload using the no-reuse strategy, which does not recycle any intermediates but also has no cost for materialization.
Afterwards, we executed the two reuse strategies: (1) materialization-based reuse where intermediate results are spilled out to a temporary table in memory, and 
(2) \system{} which reuses internal hash-tables.
We implemented all approaches (1) and (2) as well as the no-reuse strategy in \system{}.
In order to compare (1) with our reuse strategy (2), we materialize the same intermediates as \system{} does (i.e., the input of joins for which we build a hash table and outputs of aggregations).
Moreover, as described in \cite{vvreuse_13} (1) only supports exact and subsuming-reuse but neither partial nor overlapping-reuse. 

The results of this experiment are shown in Figure \ref{fig:exp:single}, where we first present the resulting total runtime with some statistics (e.g., cache sizes, hit ratio) for executing the different workloads.
In this experiment, we turned the garbage collection (GC) off. 
The effects of GC are analyzed in Section \ref{sec:eval:exp4}.

Figure \ref{fig:exp:runtime} shows the overall speed-up of both reuse strategies over the no-reuse strategy when running under different workloads. 
We see that our strategy in \system{} shows the highest speed-up for all workloads (low-, medium-, and high-reuse).
For the workload with high reuse potential \system{} achieves a speed-up of $90\%$ over the no-reuse strategy, while the materialization-based reuse strategy only achieves $40\%$. 
For the workload with low-reuse potential which simulates a user randomly browsing the data, \system{} has a performance comparable to the no-reuse strategy; i.e., it does not introduce additional overhead even if there is (almost) no reuse potential.
This is different for the materialized-reuse which introduces a slow down of $10\%$ caused by the additional materialization costs.

Figure \ref{fig:exp:stats} shows additional statistics for each workload: memory footprint, hit ratio, and the total runtime.
For the materialized-reuse strategy, we report the memory footprint for all temporary tables as well as the hit ratio per temporary table (i.e., how often a temporary table was reused).
For \system{}, we report the footprint for all cached intermediate hash-tables tables as well as the hit ratio per hash table.
The hit ratio is given as the average ratio of how often each element in the cache was re-used by a query. 

For the medium- and high-reuse case, we see that \system{} requires less memory in total while providing a higher hit ratio per cached element than the materialized-reuse strategy.
The main reason for this is that the materialized-reuse strategy only supports two out of four reuse-cases supported by \system{}.
To that end, less intermediates are reused and more new ones are added to the cache.
For the low-reuse case, we see that the hit ratio of the cache is almost $0$ in both strategies.
In this case, the memory footprint is the highest since queries just register new elements to the cache without actually reusing them.
Moreover, the memory footprint of \system{} is slightly higher than the materialized-reuse case.
The reason is that hash tables have an additional overhead (e.g., pointers for linked lists of extends) when compared to a temporary table which is essentially an array in memory without any overhead.
However, it is interesting to note that this does not have an effect on the runtime of \system{} since caching the internal hash tables does not cause any additional memory I/O compared to the no-reuse strategy.
This is different from the materialized-reuse strategy, which requires additional I/O to persist the output of operators to the memory in order to support reuse.

\subsection{Exp. 2: Efficiency of Query Optimizer}
\label{sec:eval:exp1}

In this experiment, we show the benefits of our reuse-aware optimizer.
We therefore study the runtime of 
(a) reuse on the query-level as well as 
(b) reuse on the operator-level (i.e., for reuse-aware joins and aggregations).
The main goal is to compare the performance of our cost-model based strategy with two baselines: the first baseline is \emph{never-share}, where we turn reuse in our system completely off. The second baseline is \emph{always-share}, where all operators use a greedy-heuristic to reuse the matching hash table in the cache with the highest reuse ratio. We include this strategy in order to show that greedily reusing hash tables can result in a performance that is even worse than the performance of the \textit{Never Share} strategy and to emphasize the need for a cost model that decides whether to reuse a hash table or not.\\

\vspace{-1.5ex}
\textbf{Exp. 2a - Reuse on the Query-level:} 
In this experiment, we selected a subset of seven queries from the workload with high-reuse potential.
We selected these queries, since each query represents a different type of user interaction and thus provides different reuse potentials for join and aggregation operators.
We selected the high-reuse case in order to show that the always-share baseline might result in non-optimal plans and showing that our cost-based approach finds better reuse-aware plans.

\begin{figure}
  \begin{minipage}[b]{.23\textwidth}
    \centering
    \includegraphics[width=\textwidth, height=0.6\textwidth, trim=5mm 0mm 4mm 0mm]{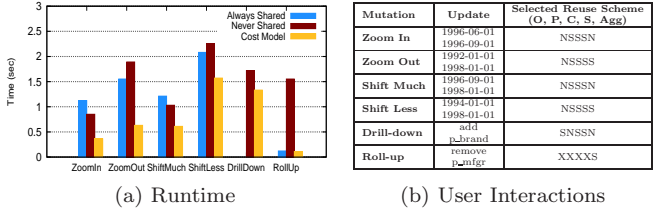}
    \vspace{-3.5ex}
    \subcaption{Runtime}
    \label{fig:exp:states}
  \end{minipage}
  \hfill
  \begin{minipage}[b]{.23\textwidth}
    \centering
    \scriptsize
    \resizebox{\columnwidth}{!}{
    \begin{tabular}{|l|c|c|}
    \hline
    {\bf Mutation}  & {{\bf Update}} & {\bf \begin{tabular}[c]{@{}c@{}}Selected Reuse Scheme\\ (O, P, C, S, Agg)\end{tabular}} \\ \hline
    {\bf Zoom In}    & \begin{tabular}[c]{@{}c@{}}1996-06-01\\ 1996-09-01\end{tabular} & NSSSN                                          \\ \hline
    {\bf Zoom Out}   & \begin{tabular}[c]{@{}c@{}}1992-01-01 \\ 1998-01-01\end{tabular}                    & NSSSS                                          \\ \hline
    {\bf Shift Much} & \begin{tabular}[c]{@{}c@{}}1996-09-01\\ 1998-01-01\end{tabular}  & NSSSN                                          \\ \hline
    {\bf Shift Less} & \begin{tabular}[c]{@{}c@{}}1994-01-01\\ 1998-01-01\end{tabular}                    & NSSSS                                          \\ \hline
    {\bf Drill-down} & \begin{tabular}[c]{@{}c@{}}add\\ p\_brand\end{tabular}                  & SNSSN                                          \\ \hline
    {\bf Roll-up}    & \begin{tabular}[c]{@{}c@{}}remove\\ p\_mfgr\end{tabular}                & XXXXS                                          \\ \hline
    \end{tabular}
    }
    \subcaption{User Interactions}
    \label{exp:table:states}    
  \end{minipage}
  \\
  \vspace{-1.5ex}
  \caption{Reuse on the Query-Level}
  \vspace{-6.5ex}
\end{figure}

The first query of the sequence we picked is a 5-way SPJA query over the tables  \texttt{Lineitem}, \texttt{Orders}, \texttt{Part}, \texttt{Customer}, and \texttt{Supplier}. 
The details of the six follow-up queries are summarized in Table \ref{exp:table:states}.
The first column of this table lists the type of user interaction that was applied . 
The second column shows the difference of each query to its predecessor:
The first four follow-up queries modify the selection predicate on the attribute \texttt{o\_orderdate}.
The last two queries modify the group-by keys.

For running this experiment, we executed all seven queries sequentially over the TPC-H database using our reuse strategy as well as using the two baselines (never-share and always-share).
The first query populates the cache with five hash tables in total: four resulting from the joins and one from the aggregation. 
The results for the six follow-up queries (that are candidates for reuse) are shown in Figure \ref{fig:exp:states}.
In this figure, we see that the \emph{Cost Model} strategy, which is based on our optimizer, outperforms the two other baselines since it always picks the optimal reuse strategy.
In the best case (i.e., the \emph{RollUp} follow-up query), the speed up factor is about two orders of magnitude better than \emph{never-share}. 
The reason is that the cached aggregation hash table is sufficient to execute the \emph{RollUp} query (i.e., no missing tuples need to be added and thus no joins need to be executed at all).
For the \emph{Drill Down} query, we could not execute the \textit{Always Share} strategy since the \emph{p\_brand} attribute was never included in the corresponding hash table in previous executions and thus that hash table is not reusable.

The last column of Table \ref{exp:table:states} shows the detailed decisions of our optimizer (i.e., for the \textit{Cost Model} strategy) for all operators of the six follow-up queries, which explain our performance results in Figure \ref{fig:exp:states}.
The string in this column uses one character to encode the decision for each operator (join and aggregation).
The operators from left to right are shown in the header of the last column:
For example, the $O$ character represents the hash table created by the build phase of a join that scanned the \texttt{Orders} relation. 
The other characters represent the hash tables created by the build phase that scanned the \texttt{Part}, \texttt{Customer}, and \texttt{Supplier} tables.
\emph{Agg} represents the aggregation operator that is executed on top of all joins. 
The characters encode the following decision:
$N\ (Not\ Shared)$ states that a new hash table was created for the operator whereas $S\ (Shared)$ states that the existing hash table was reused. 
Moreover, $X$ defines that this operator was not need to be executed at all for the given query. 
For instance, this case occurs for the \textit{Roll Up} operation, where the new query just needs to read the cached aggregation hash table. 
\\

\vspace{-1.5ex}
\textbf{Exp. 2b - Reuse on the Operator-level (RHJ):} In this experiment, we show the efficiency of our optimizer for the reuse-aware hash-join (RHJ). 
For showing the efficiency, we directly execute the reuse-aware hash-join operators on two input synthetic tables.
The table for the building phase was $16MB$ in size and the table for the probing phase had $10\times$ the size of the table for the build phase.

In order to show the efficiency of our optimizer for RHJ, after adding a candidate hash table of $16MB$ to the hash table cache we executed multiple runs with different contribution ratios from $100\%$ to $0\%$.
$100\%$ contribution-ratio means that the RHJ can reuse all tuples in the cached hash table and does not need to post-filter any tuples after probing; whereas $0\%$ contribution-ratio means that the RHJ can not reuse any tuples in the cached hash table.
Moreover, $0\%$ contribution-ratio means there is $100\%$ overhead in the reused hash table (i.e., all tuples must be post-filtered) due to the fact that for all contribution-ratios we keep the size of the cached hash-table the same.

\begin{figure}
  \begin{minipage}[b]{.23\textwidth}
    \centering
    \includegraphics[width=\textwidth, trim=5mm 0mm 4mm 0mm]{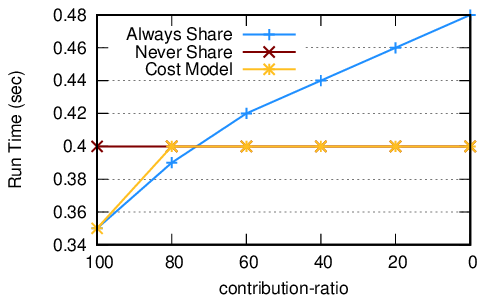}
    \vspace{-2.5ex}
     \subcaption{ Reuse-Aware Hash-Join }
    \label{fig:exp:join_16_16}
  \end{minipage}
  \hfill  
    \begin{minipage}[b]{.23\textwidth}
    \centering
    \includegraphics[width=\textwidth, trim=5mm 0mm 4mm 0mm]{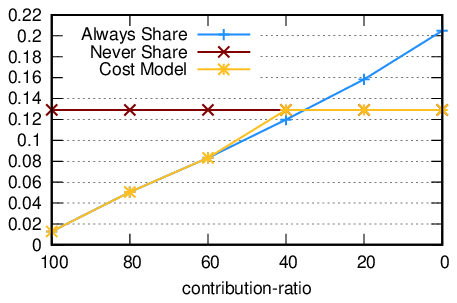}
     \vspace{-2.5ex}
     \subcaption{ Reuse-Aware Hash-Agg. }
    \label{fig:exp:agg}
    \end{minipage}
  \\
  \vspace{-1.5ex}
  \caption{Reuse on the Operator-Level}
  \label{fig:join_exp}
  \vspace{-4.5ex}
\end{figure}

Same as in the previous experiment, here we compare our \textit{Cost Model} based strategy against the \textit{Never Share} (i.e., a traditional hash-join) and the \textit{Always Share} strategy which always picks the cached hash table for reuse.
Figure \ref{fig:exp:join_16_16} shows the results.
We see that the \textit{Never Share} strategy pays a constant price since it never reuses the hash table.
Moreover, the costs for the \text{Always Share} strategy are constantly increasing since more and more missing tuples need to be added to the reused hash table (if the contribution-ratio decreases).
At approx. $70\%$ contribution-ratio, the \text{Always Share} gets more expensive than the \textit{Never Share} strategy due to the overhead incurred in the cached hash table (i.e., tuples in the hash table that are not required by the RHJ).
As an important result, we see that our \text{Cost Model} always picks the best strategy with the minimal cost: for a contribution-ratio from $100\%$ to $70\%$ it reuses the cached hash table and below $70\%$ it decides to create a new hash table since the total runtime costs are cheaper when not reusing the candidate hash table in the cache.\\

\vspace{-1.5ex}
\textbf{Exp. 2c - Reuse on the Operator-level (RHA):} 
In this experiment, we show the effect of reusing hash tables for reuse-aware hash-aggregates (RHAs).
We again varied the contribution-ratio of the cached hash table as in the experiment before.
Figure \ref{fig:exp:agg} shows that our cost model still picks the best strategy with the minimal cost.

\subsection{Exp. 3: Accuracy of Query Optimizer}
\label{sec:eval:exp2}

\begin{figure}
\centering
\includegraphics[width=.4\textwidth]{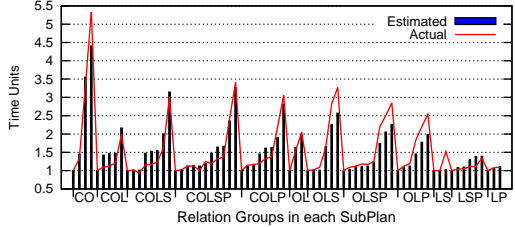}
\vspace{0.5ex}
\caption{Accuracy of Cost Models}
\vspace{-6.5ex}
\label{fig:exp:enum}
\end{figure}

As described in Section \ref{sec:single:pe}, the plan enumeration algorithm is one of the core elements of \system{} to select a reuse-aware plan with minimal runtime for a given set of cached hash tables.
In this experiment, we validate the accuracy of the cost estimation component of our optimzer (i.e., the \texttt{cost} function used in Algorithm \ref{alg:planenum}).

For this experiment, we execute the workload described in Section \ref{sec:eval:exp0} with medium-reuse potential.
In order to analyze the accuracy of our cost estimation, we select one of the $5$-way join queries over the tables \texttt{Lineitem}, \texttt{Orders}, \texttt{Part}, \texttt{Customer}, and \texttt{Supplier} from this workload and analyze the estimated and actual cost of the optimizer. 
We selected this query since it is a complex query with multiple joins and the optimal reuse-aware plan contains both cases: operators that reuse a cached hash table and other operators that create a new hash table.
In order to analyze the accuracy of our cost estimates, we compare the estimated and the actual cost for each enumerated sub-plan of this query.
Figure \ref{fig:exp:enum} shows the results.

As a general observation, we can see that our cost models are accurate since the actual and estimated costs follow the same trend.
To better understand the results, we clustered the costs into groups that represent equivalent sub-plans (i.e., one group represents sub-plans over the same partition of the join graph).
For example, the group \texttt{CO} represents the enumerated join plans over the two tables \texttt{Customer} and \texttt{Orders} for all hash tables in the cache.
Moreover, to better compare the actual and estimated costs, we are using normalized costs (called time units). 
For normalization, the lowest cost per group uses the cost of $1$.

As discussed in Section \ref{sec:single:pe}, plan enumeration works incrementally and picks the cheapest sub-plan per group and composes the complete plan based on these optimal sub-plans.
Thus, the quality of the optimizer depends only on the fact that whether or not it finds the cheapest plan per group.
In order to see this, if the optimizer finds the cheapest plan per group, the normalized costs are sufficient (i.e., the absolute costs do not matter for this decision).
Figure \ref{fig:exp:enum} orders the sub-plans per group by their actual costs.
As shown in Figure \ref{fig:exp:enum}, the estimated costs follow the same trend as the actual costs.
Even more important is that the first plan per group, which has the lowest actual cost resulting from the ordering, always has the lowest estimated cost as well.
To that end, our optimizer is able the find the most optimal sub-plan per group for the query.

\subsection{Exp. 4: Multi-Query Reuse}
\label{sec:eval:exp3}

In this section, we present the evaluation results for the query-batch interface as explained in Section \ref{sec:multi}.
In order to generate the batches of queries, we group the query trace of $64$ queries of the workload with medium-reuse potential of the experiment in Section \ref{sec:eval:exp0} into smaller sets of $4$, $8$, and $16$ queries.
In order to populate the cache, we first executed one batch of the given size (e.g., $4$ queries) and afterwards executed $10$ randomly selected batches of the same size and report the average run time for one batch as a result. 

Moreover, in order to show the effect of reuse in shared plans, we executed the same sequence of batches using different modes:
The first mode (\emph{single-query plan, wo reuse}) is the traditional database execution mode where queries are executed individually that do not reuse any cached hash tables. 
The second mode (\emph{single-query plan, w reuse}) executes all queries individually as well, but in a set up where reuse of intermediate hash tables using our cost-model is enabled.
The last mode (\emph{shared plan, w reuse}) represents the mode where we use our reuse-aware shared plans as introduced in Section \ref{sec:multi} to execute the batch.

\begin{figure}
\centering
\includegraphics[width=.3\textwidth, trim=4mm 0mm 5mm 0mm]{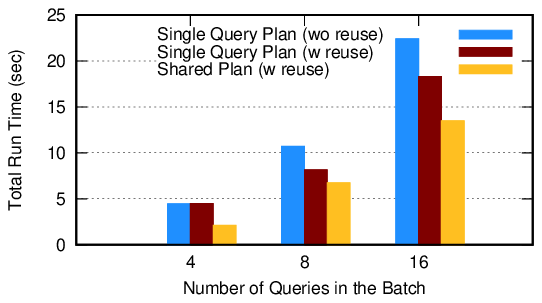}
\caption{Batch Execution  in \system{}}
\vspace{-4.5ex}
\label{fig:exp:batch}
\end{figure}

Figure \ref{fig:exp:batch} shows the total runtime summed up  for all queries in a batch of a given size.
As expected, the first mode (\emph{single-query plan, wo reuse}) has the highest total runtime for all batch sizes.
The second mode (\emph{single-query plan, w reuse}) reduces the total runtime on average by approx. $20\%$ to run all queries in the batch, which purely results from reusing intermediates.
Finally, the execution of the (\emph{shared plan, w reuse}) mode results in the lowest total runtime, which is on average approximately $40\%$ lower than for the first mode (\emph{single-query plan, wo reuse}).
The additional reduction in runtime compared to the \emph{single-query plan without reuse} mode result from the effect of shared reuse-aware plans. Our plan enumeration algorithm of the query-batch interface in \system{} creates on average, $2$ shared reuse-aware plan to execute the batch of size $4$, $3$ plans for the batch of size $8$, and $4$ plans for the batch of size $16$.

\subsection{Exp. 5: Effects of Garbage Collection}
\label{sec:eval:exp4}

In this experiment we show the effect of garbage collection on the performance of \system{}. 
We again used the workloads as described in Section \ref{sec:eval:exp0} and executed the complete trace using two modes:
The first mode (\emph{wo GC}) represents the case where we execute all queries using the query-at-a-time interface with reuse, however, no garbage collector was active; i.e., \system{} used as much memory as needed to cache all hash tables.
For the second mode (\emph{with GC}), we additionally activated the garbage collector (GC).
For the hash table cache, we used $20\%$ of the memory that would be required to store all hash tables.

As a result, we measured the additional runtime overhead that was caused by the effects of the garbage collector (i.e., monitoring the size of all caches hash tables, evicting and reloading evicted hash tables).
Compared to \system{} without GC, our experiment shows that \system{} with GC introduces approximately only a $10\%$ higher overhead for the medium- and high-reuse case.
For the high-reuse case this is negligible when looking at the performance gains of \system{} over a DBMS without any reuse (as we have shown in the experiments before).
For the medium-reuse case, \system{} can still achieve a performance speed-up of $10\%$ over the no-reuse case.
Note, however, that when increasing the cache size to $50\%$ of the total memory required to cache all hash tables, the overhead of garbage collection drops down to $5\%$.
Most interestingly, for the low-reuse workload, GC causes almost no overhead  since intermediate hash-tables are anyway almost never reused.

\vspace{-1.5ex}
\section{Related Work}
\label{sec:related}

\textbf{Reuse of Intermediates:}  
In order to better support user sessions in DBMSs, various techniques have been developed in the past to reuse intermediates \cite{cod2001,monetdb2009,vvreuse_13}.
All these techniques typically require that results of individual operators are materialized into temporary tables.
This is very different from \system{}, which revisits "reuse" in the context of modern main memory DBMSs and promotes to leverage internal hash tables that are materialized by pipeline breakers and thus does not add any additional materialization cost to query execution.

In the following, we discuss further differences when comparing these techniques to the ideas of \system{}.
\cite{cod2001} introduces an optimizer to select which intermediates should be reused.
Different from \system{}, the cost models are rather coarse-grained and centered around the I/O benefits in disk-based DBMS. To that end, their cost models do not take the peculiarities of hash tables as well as hardware-dependent parameters such CPU caches into account.
In \cite{monetdb2009}, the authors integrate reuse techniques into MonetDB, that implements an operator-at-a-time execution model which anyway relies on full materialization of all intermediate results and thus does not need to tackle the issues that result form additional materialization cost as in pipelined databases.
\cite{vvreuse_13} extends the work of \cite{monetdb2009} for pipelined databases and integrates the ideas into Vectorwise. 
In this paper, the authors introduce a cache with lineage which is similar to the ideas of the hash table manager in \system{}. 
A major difference is, however, that in both cases intermediate results of operators are reused and not internal data structures of operators as we suggest in \system{}.
Moreover, compared to all the approaches mentioned before \cite{cod2001,monetdb2009,vvreuse_13}, our work also supports reuse-cases for partial- and overlapping reuse and most importantly introduces a reuse-aware optimizer.

Another area where reuse of intermediates was analyzed is in the context of Hadoop.
ReStore \cite{redoop2014} is able to reuse the output of whole MapReduce jobs that are part of a workflow implemented in PigLatin.
Moreover, it additionally materializes the output individual operators that are executed within a MapReduce job.
Since Restore is based on Hadoop and not tailored towards reuse in main memory systems, it makes their reuse techniques fundamentally different from those presented in \system{}.

Finally, buffer pools and query caches in database systems \cite{querycache1994,semcache1996} serve as a cache for frequently accessed data. 
However, the main purpose of buffer pools and query caches is to speed-up the access to base data (in case of the database buffer) or the final query result (in case of query caches) but not to reuse intermediates.\\

\vspace{-1.5ex}
\textbf{Materialized Views:} 
Reusing results has also been the main motivation of materialized views \cite{matview2001}.
Again, the main difference of materialized views and  \system{} is that reuse is for materialized views is externalized as an additional "table" rather than leveraging internal data structures that are produced by query processing.
Moreover, our reuse-aware optimizer implements cost-models that target the reuse of internal data structures and introduces benefit-oriented optimizations, both aspects that are not covered by traditional  optimizers that rewrite queries for materialized views.
\\

\vspace{-1.5ex}
\textbf{Automatic Physical Schema Design:} 
Another line related to our work are techniques for online physical schema tuning \cite{online_tuning07}.
The main goal of this work is to create additional database objects such as indexes or materialized views (discussed before) without involving a database administrator.
Adaptive indexing \cite{adaptive_idx11,partial_idx89} also falls into this category and suggests to create indexes partially as a side effect of query processing.
However, again these techniques do not consider internal data structures for reuse but externalize their decision by creating additional (partial) indexes, views, etc.
\\

\vspace{-1.5ex}
\textbf{Multi-Query-Optimization:} Another area of related work is Multi-Query-Optimization (MQO) \cite{mqo}.
The main idea of MQO is to identify common sub-expressions of a set of queries that are active in a DBMS at the same time.
In order to save resources, common sub-expressions are only executed once.
One problem of MQO is that in most workloads, common sub-expressions are a rather rare case.
Therefore, MQO is typically used to optimize OLAP workloads over a star schema where the chance of common sub-expressions is higher since most queries join the dimension tables with the same fact table.
All ideas in MQO are orthogonal to the reuse ideas presented in this paper; i.e., reuse of hash tables can be integrated into plans created by MQO techniques.
\\

\vspace{-1.5ex}
\textbf{Work-Sharing:} Work-sharing systems \cite{shredscanibm2008,sharedmonet2007,crescando2009,qpipe2006,cjoin2009,sharedb2014} have similar goals as MQO since they also process multiple queries at a time by sharing work.
However, different from MQO they do not require to identify the very same sub-expression to share work.
One of the techniques for work-sharing is the shared (or cooperative) scan operator \cite{shredscanibm2008,crescando2009,sharedmonet2007}. 
The idea of shared scans is that the scan operation can be shared by queries even if queries use different selection predicates.
Other systems such as QPipe \cite{qpipe2006}, CJoin \cite{cjoin2009}, SharedDB \cite{sharedb2014} extend the idea of work-sharing to other operators such as joins and aggregations. 
All these ideas for work-sharing are again orthogonal to the reuse ideas presented in this paper.
In this paper, we actually extended the ideas of \cite{sharedb2014} to integrate reuse into shared-plans.
\vspace{-1.5ex}
\section{Conclusions and Future Work}
\label{sec:concl}

Salient characteristics of modern main memory DBMSs and interactive analytical workloads require a critical rethinking of reuse in query processing. Our solution, called \system{}, focuses on the reuse of hash tables populated with intermediate query results. We avoid additional materialization costs by leveraging hash tables that are already materialized at pipeline breakers. We also do not incur the overhead of casting hash tables to relations and vice versa by treating hash tables as native units of reuse. Our reuse-aware optimizer can accurately model hash table usage and its impact in query performance, leading to highly profitable reuse choices that offer up to $100\times$ performance improvement over the no-reuse baseline for realistic workloads.
  
We plan to extend the ideas presented in this paper to other data structures (e.g., trees) as well as concurrent reuse.


\newpage
\begin{scriptsize}
\bibliographystyle{abbrv}
\bibliography{bib}

\end{scriptsize}

\end{document}